\documentclass{SciPost}

\hypersetup{
    colorlinks,
    linkcolor={red!50!black},
    citecolor={blue!50!black},
    urlcolor={blue!80!black} 
}

\usepackage[bitstream-charter]{mathdesign}

\DeclareSymbolFont{usualmathcal}{OMS}{cmsy}{m}{n}
\DeclareSymbolFontAlphabet{\mathcal}{usualmathcal}

\fancypagestyle{SPstyle}{
\fancyhf{}
\lhead{\colorbox{scipostdeepblue}{\bf \color{white} ~SciPost Physics Proceedings }}
\rhead{{\bf \color{scipostdeepblue} ~Submission }}

\fancyfoot[C]{\textbf{\thepage}}
}

\usepackage{amsmath}
\usepackage{bm}
\usepackage{dsfont}
\usepackage{url}

\newcommand{\vtheta}{{\vec{\theta}}}
\newcommand{\bra}[1]{\langle #1 |}
\newcommand{\ket}[1]{| #1 \rangle}
\newcommand{\braket}[1]{\langle #1 | #1 \rangle}
\newcommand{\ev}[2]{\langle #2 | #1 | #2 \rangle}

\begin{document}

\pagestyle{SPstyle}

\begin{center}{\Large \textbf{\color{scipostdeepblue}{
Quantum Dynamics with Time-Dependent Neural Quantum States
}}}\end{center}

\vspace{-1em} 

\begin{center}\textbf{
Alejandro Romero-Ros\textsuperscript{1, 2, $\dagger$},
Javier Rozalén Sarmiento\textsuperscript{1, 2} and
Arnau Rios\textsuperscript{1, 2}
}\end{center}

\begin{center}
{\bf 1} Departament de Física Quàntica i Astrofísica (FQA), Universitat de Barcelona (UB),  c. Martí i Franqués, 1, 08028 Barcelona, Spain
\\
{\bf 2} Institut de Ciències del Cosmos (ICCUB), Universitat de Barcelona (UB), c. Martí i Franqués, 1, 08028 Barcelona, Spain
\\[\baselineskip]
$\dagger$ \href{mailto:alejandro.romero.ros@fqa.ub.edu}{\small alejandro.romero.ros@fqa.ub.edu}
\end{center}

\definecolor{palegray}{gray}{0.95}
\begin{center}
\colorbox{palegray}{
  \begin{tabular}{rr}
  \begin{minipage}{0.37\textwidth}
    \includegraphics[width=60mm]{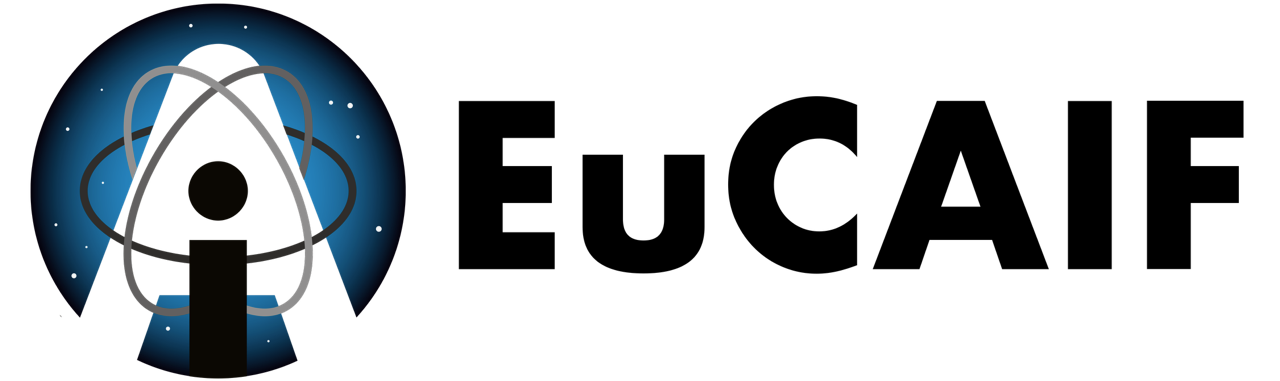}
  \end{minipage}
  &
  \begin{minipage}{0.5\textwidth}
    \vspace{5pt}
    \vspace{0.5\baselineskip} 
    \begin{center} \hspace{5pt}
    {\it The 2nd European AI for Fundamental \\Physics Conference (EuCAIFCon2025)} \\
    {\it Cagliari, Sardinia, 16-20 June 2025
    }
    \vspace{0.5\baselineskip} 
    \vspace{5pt}
    \end{center}
    
  \end{minipage}
\end{tabular}
}
\end{center}

\section*{\color{scipostdeepblue}{Abstract}}
\textbf{\boldmath{%
We present proof-of-principle time-dependent neural quantum state (NQS) simulations to illustrate the ability of this approach to effectively capture key aspects of quantum dynamics in the continuum.
NQS leverage the parameterization of the wave function with neural-network architectures.
Here, we put NQS to the test by solving the quantum harmonic oscillator.
We obtain the ground state and perform coherent state and breathing mode dynamics.
Our results are benchmarked against analytical solutions, showcasing an excellent agreement.
}}

\vspace{\baselineskip}

\noindent\textcolor{white!90!black}{%
\fbox{\parbox{0.975\linewidth}{%
\textcolor{white!40!black}{\begin{tabular}{lr}%
  \begin{minipage}{0.6\textwidth}%
    {\small Copyright attribution to authors. \newline 
    This work is a submission to SciPost Phys. Proc. \newline
    License information to appear upon publication. \newline
    Publication information to appear upon publication.}
  \end{minipage} & \begin{minipage}{0.4\textwidth}
    {\small Received Date \newline Accepted Date \newline Published Date}%
  \end{minipage}
\end{tabular}}
}}
}



\section{Introduction}
\label{sec:intro}

In recent years, the use of neural networks (NN) in machine learning applications has revolutionized almost every field of research, reducing both computational costs and time.
Here, we discuss Neural Quantum States (NQS)~\cite{Medvidovic2024} as a promising scalable framework for solving quantum many-body problems where the exponential growth with particle number becomes a challenge, ranging from spin systems~\cite{Carleo2017} to quantum chemistry~\cite{Nys2024}.

A NQS consists of a wavefunction ansatz parametrized by the weights and biases of a NN.
The underlying NN architecture can be designed to capture specific symmetries and physical constraints of the system under study~\cite{Keeble2020,Keeble2023,RozalenSarmiento2024}.
Additionally, the time-dependent variational principle allows one to derive specific equations of motion for the NN parameters to compute the evolution of the wavefunction~\cite{Hackl2020}.

As a proof of principle, in this work we employ a time-dependent NQS to simulate single-particle dynamics in a one-dimensional harmonic trap. 
In particular, we present two case studies:
the dynamics of a coherent state and the excitation of a breathing mode (monopole oscillations).
To benchmark the results, we compare them with exact analytic solutions~\cite{Zwiebach2022}.

\section{Time-dependent Variational Principle}
\label{sec:equations_of_motion}

  A wavefunction ansatz $\ket{\Psi_\vtheta}$ parametrized by a set $\vec{\theta}=\{\theta_1,\dots,\theta_M\}$ of $M$ complex parameters defines a variational manifold $\mathcal{M}$ spanned by all such states.
  For a Hamiltonian $\hat{\mathcal{H}}$, the time evolution of $\ket{\Psi_\vtheta}$ over a small time-step $\delta t$ is governed by the unitary time operator, $\ket{\Psi_{\vtheta}(t+\delta t)} = e^{-i\hat{\mathcal{H}}\delta t/\hbar}\ket{\Psi_{\vtheta}(t)}$.
   
  If we allow the parameters to be time-dependent, $\vtheta\equiv\vtheta(t)$ with $\delta\vtheta=\dot{\vtheta}\delta t$, and provided that $\mathcal{M}$ is large enough, the same dynamics can be captured by a parameter trajectory, i.e., $\ket{\Psi_{\vtheta}(t+\delta t)} \approx \ket{\Psi_{\vtheta+\delta \vtheta}}$.
  Now, the equations of motion can be obtained applying the Dirac-Frenkel variational principle~\cite{Hackl2020}.
  This principle is equivalent to solving the Euler-Lagrange equations (EL),
  $   \frac{\partial\mathcal{L}}{\partial\theta_\mu} - \frac{d}{dt}\frac{\partial\mathcal{L}}{\partial\dot{\theta}_\mu} = 0 $
  and 
  $ \frac{\partial\mathcal{L}}{\partial\theta^*_\nu} - \frac{d}{dt}\frac{\partial\mathcal{L}}{\partial\dot{\theta}^*_\nu} = 0$,
  where 
  \begin{equation}
  \label{eq:lagrangian}
      \mathcal{L}\left(\vtheta,\dot{\vtheta},\vtheta^*,\dot{\vtheta}^*\right) 
      = \frac{\bra{\Psi}i\hbar\frac{d}{dt} - \hat{\mathcal{H}}\ket{\Psi}}{\braket{\Psi}}
      = i\hbar\sum_{\mu=1}^m\frac{\ev{\partial_\mu}{\Psi}}{\braket{\Psi}}\dot{\theta_\mu}
      + i\hbar\sum_{\nu=1}^m\frac{\ev{\partial_\nu}{\Psi}}{\braket{\Psi}}\dot{\theta^*_\nu} - \frac{\ev{\hat{\mathcal{H}}}{\Psi}}{\braket{\Psi}}
  \end{equation}
  is the Lagrangian of the system expressed as a function of $\vtheta$.
  Here, we drop the dependency on $t$ and $\vtheta$ of $| \Psi \rangle$ and use $\frac{\partial}{\partial\theta_\mu}=\partial_\mu$ and $\frac{\partial}{\partial\theta^*_\nu}=\partial_\nu$, for clarity. 

  Assuming a holomorphic wavefunction, i.e., $\frac{\partial\ket{\Psi}}{\partial\theta^*_\nu}=0$, we obtain the following equation of motion for the parameters:
  \begin{align}
  \label{eq:eom_holomorphic}
      \bm{S}\dot{\vtheta} = -\frac{i}{\hbar}\vec{F}^*
      \,,
  \end{align}
  with 
  \begin{align}
  \label{eq:S}
      S_{\nu\mu} &= 
          \frac{\partial_{\nu}\bra{\Psi}\partial_{\mu}\ket{\Psi}}{\braket{\Psi}}
      -
          \frac{\partial_{\nu}\braket{\Psi}}{\braket{\Psi}}\frac{\ev{\partial_\mu}{\Psi}}{\braket{\Psi}} 
          \,,
      \\
  \label{eq:F}
      F^*_{\nu} &=
      \frac{\partial_{\nu}\bra{\Psi}\hat{\mathcal{H}}\ket{\Psi}}{\braket{\Psi}} 
      - \frac{\partial_{\nu}\braket{\Psi}}{\braket{\Psi}}\frac{\ev{\hat{\mathcal{H}}}{\Psi}}{\braket{\Psi}}
          \,,
  \end{align}
  where $\bm{S}$ is the so-called Quantum Geometric Tensor, and $\vec{F}$ are the variational forces.

\section{Methods}
\label{sec:methods}

  We employ a NQS parametrized by a NN with one input and one output neuron, and a hidden layer of 5 neurons.
  All neurons use a sigmoid activation function.
  Including biases, the NQS has $M=16$ complex-valued parameters, initialized with Xavier initialization~\cite{Glorot2010}.   
  
  The simulations take place in a finite grid $x_i\in[-8,8]$ with $n=100$ points.  %
  The NN input is a grid point $x_i$.
  The output, $f(x_i)$, relates to the wavefunction through $\log\Psi(x_i)=e^{f(x_i)}$, which we found to improve numerical stability.
  
  The dynamics of the system are computed solving Eq.~\eqref{eq:eom_holomorphic}.
  At every step, the wavefunction is determined by $\vtheta$.
  Then, it is used to compute $\bm{S}$, $\vec{F}$ and $\dot{\vtheta}$, and to update $\vtheta$ accordingly.
  Since $\bm{S}$ is generally ill-conditioned, and thus its inversion is computationally unstable, we regularize it with a diagonal shift $\bm{S}\leftarrow\bm{S}+\lambda\mathds{1}$, with $\lambda \in \mathds{C}$ and $|\lambda| \ll 1$. 
  
  Ground states are obtained similarly, but using imaginary time propagation, with $t \to it$ in Eq.~\eqref{eq:lagrangian}, which yields $\bm{S}\dot{\vtheta} = -\hbar^{-1}\vec{F}^*$.
  Then, convergence is reached when $|\delta E| \leq 10^{-8}$ for 10 consecutive steps, where $E=\frac{\ev{\hat{\mathcal{H}}}{\Psi}}{\braket{\Psi}}$ is the energy.

  The NQS is built using PyTorch~\cite{Paszke2019} and the equations of motion~\eqref{sec:equations_of_motion} are integrated using a 4th-order Runge-Kutta scheme with $\delta t = 0.1$, $\lambda=10^{-4}$ for ground state search and $\lambda=i10^{-6}$ for real-time evolution. 
  This setup computes the ground state in $\sim$1.5~s and dynamics up to $t=50$ in $\sim$7.5~s on a Dell Precision 3660.
  The code can be found in Ref.~\cite{Romero-Ros2025}.

\section{Results}
\label{sec:results}
  We benchmark the time-dependent NQS by simulating the dynamics of a one-dimensional harmonic oscillator (HO).
  The Hamiltonian in HO units\footnote{
    Energy, length and time are expressed in terms of $\hbar\omega_0$, $\sqrt{\hbar/m\omega_0}$ and $\omega_0^{-1}$, respectively, with $\hbar$ being the reduced Planck's constant, $m$ the particle's mass, and $\omega_0$ the characteristic trapping frequency.
    } reads
	\begin{equation}
    \label{eq:hamiltonian}
		  \hat{\mathcal{H}} = -\frac{1}{2}\partial_x^2 + \frac{1}{2}\omega^2(\hat{x}-x_0)^2
      \,,
	\end{equation}
  with frequency $\omega$ and center $x_0$.  

\begin{figure}[t!]
    \centering
    \includegraphics[width=\columnwidth]{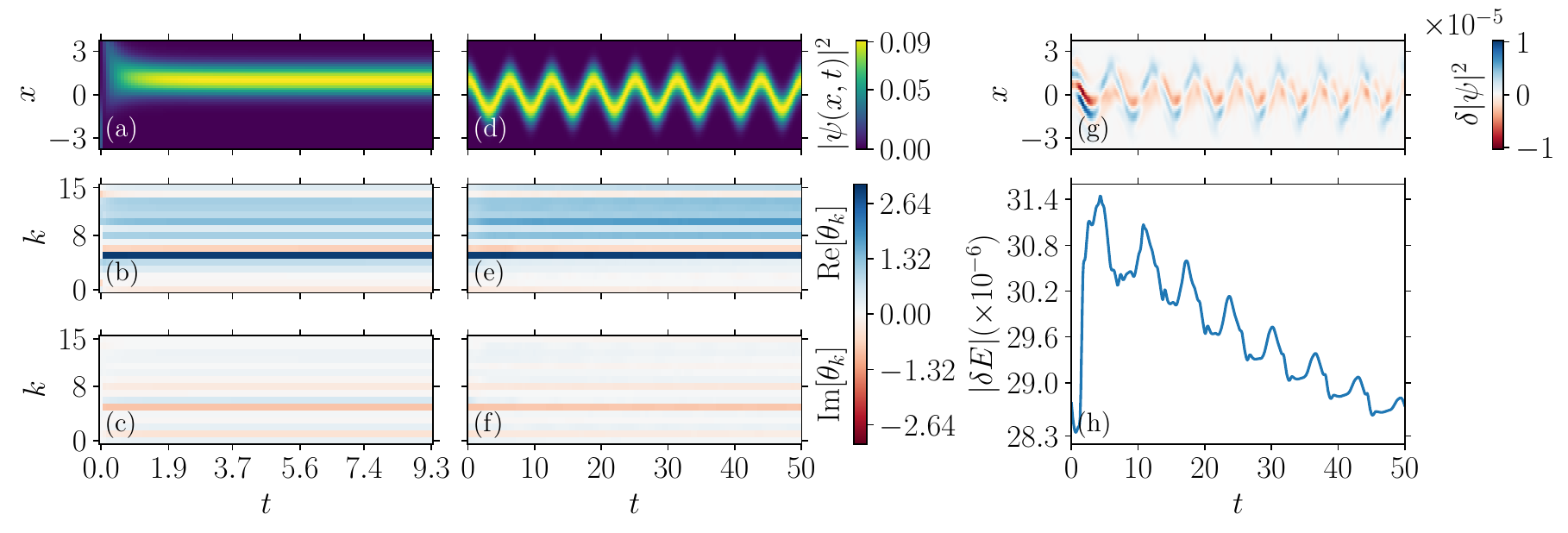}
    \caption{
      \textbf{Coherent state dynamics.} 
      (a-c): Ground state search for a displaced trap ($x_0=1$). 
      (a) Evolution of the probability density $|\psi(x,t)|^2$. 
      (b, c) Evolution of the real and imaginary parts, respectively, of the NN parameters, $\theta_k$. 
      (d-f): Dynamics after a trap displacement to $x_0=0$. 
      (d) Time evolution of the probability density. 
      (e, f) Time evolution of the real and imaginary parts of $\theta_k$, respectively. 
      (g) Error in the probability density, $\delta|\psi|^2=|\psi_{\textrm{NQS}}|^2 - |\psi_{\textrm{exact}}|^2$. (h) Deviation of the energy from its initial value, $\delta E = |E(t) - E(0)|$.
    }
    \label{fig:oscillations}
\end{figure}

  \paragraph{Coherent state dynamics}
  To prepare the initial state, we first obtain the ground state with $\omega=1$ and $x_0=1$. 
  Fig.~\ref{fig:oscillations}(a) shows the probability density converging to the expected Gaussian profile of a coherent state, while Figs.~\ref{fig:oscillations}(b) and \ref{fig:oscillations}(c) show the convergence of the real and imaginary parts of the parameters, respectively.

  To initiate the dynamics, we instantaneously shift the confinement to $x_0=0$.
  The resulting oscillations of the wavepacket are shown in Fig.~\ref{fig:oscillations}(d).
  Overall, during the dynamics the parameters vary less than 10\%, and thus no visible change is appreciated in Figs.~\ref{fig:oscillations}(e) and \ref{fig:oscillations}(f).
  Nevertheless, the NQS accurately reproduces the analytic solution, as demonstrated by the probability density error in Fig.~\ref{fig:oscillations}(g) and by the conservation of energy in Fig.~\ref{fig:oscillations}(h), with deviations up to $10^{-6}$.

\begin{figure}[t!]
    \centering
    \includegraphics[width=\columnwidth]{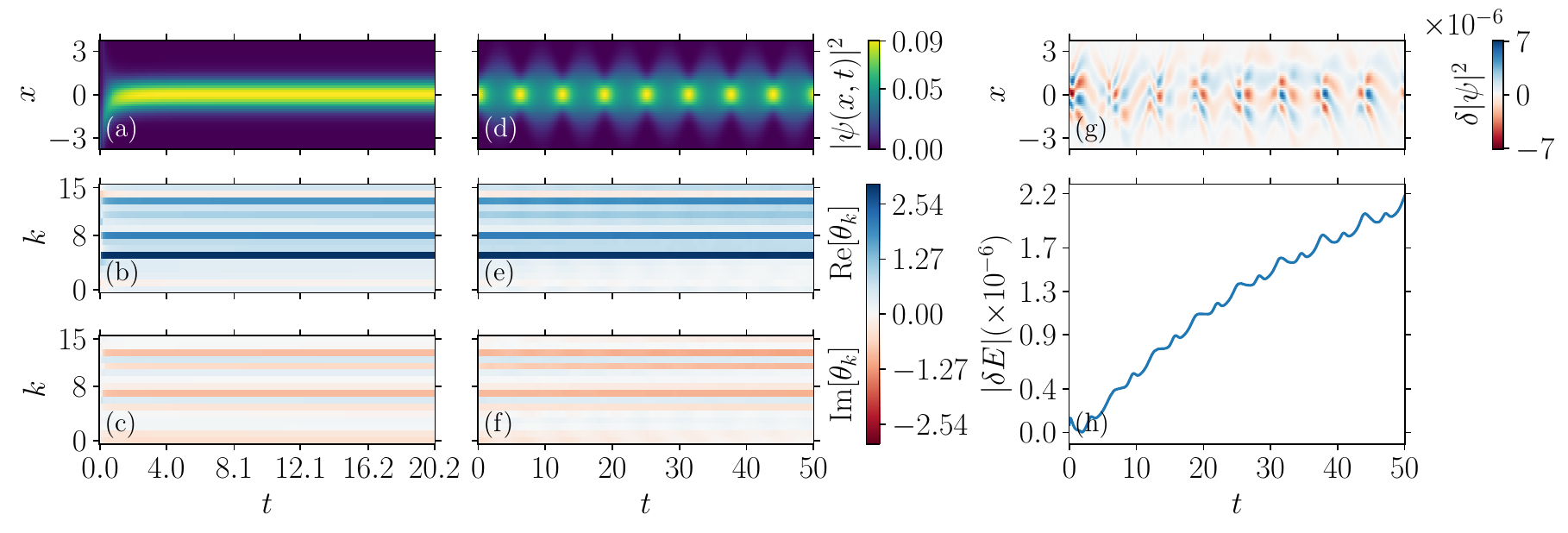}
    \caption{
      \textbf{Breathing mode dynamics.} 
      The layout is identical to Fig. \ref{fig:oscillations}.
      Left column (a-c): Ground state search for a centered trap ($x_0=0$).
      Middle column (d-f): Dynamics after a frequency quench from $\omega=1$ to $\omega=0.5$.
      (g) Error in the probability density.
      (h) Energy deviation.
    }
    \label{fig:breathing}
\end{figure}

  \paragraph{Breathing mode dynamics}
  As before, we obtain the ground state for $\omega=1$ but $x_0=0$. 
  Figs.~\ref{fig:breathing}(a)--(c) show the convergence of the probability density and the parameters.
  While the parameters differ from the previous case, some (e.g. $\theta_5$) contribute similarly, reflecting the common underlying geometry of the problem.

  To excite the breathing mode, we quench the frequency to $\omega=0.5$.
  Now, the dynamics in Fig.~\ref{fig:breathing}(d) shows the oscillations of the wavepacket's width.
  The parameter changes remain again below 10\%, hence not visible in Figs.~\ref{fig:breathing}(e) and \ref{fig:breathing}(f).
  In this case, the agreement with the analytic results is still excellent, as shown in Fig.~\ref{fig:breathing}(g) and the conservation of energy in Fig.~\ref{fig:breathing}(h), again with deviations only up to $10^{-6}$.

\section{Conclusion}
\label{sec:conclusions}
  In this work we have provided proof-of-principle simulations of a continuous one-dimensional quantum system using a time-dependent NQS.
  By benchmarking our results against the analytical solutions of the HO, we showed that with a few more than a dozen parameters a NQS can accurately reproduce both coherent-state oscillations and breathing-mode dynamics.
  In both cases, the dynamics agrees remarkably well with the exact solution throughout the evolution.  

  An important finding is that the network parameters remain stable, varying by less than 10\% during the dynamics. 
  This suggests that the NQS representation is not only expressive enough to encode the relevant wavefunctions but also robust against dynamical and, more importantly, numerical instabilities. 
  Also, the similarity in parameter relevance across distinct initializations suggests an underlying geometric structure shared by the states. 

  These results evidence the viability of time-dependent NQS as a practical variational tool for quantum dynamics.
  We expect their scalability and flexibility to become a promising framework for simulating nonequilibrium quantum many-body physics, and pave the way for higher-dimensional particle-interacting systems where exact solutions are unavailable.

\section*{Acknowledgements}
We acknowledge MCIN/AEI/10.13039/501100011033 from Grants No. PID2023-147112NB-C22; No. CNS2022-135529 through the “European Union NextGenerationEU/PRTR”; RYC2018-026072 through the “Ramón y Cajal” program funded by FSE “El FSE invierte en tu futuro”; No. CEX2024-001451-M through the “Unit of Excellence María de Maeztu 2025-2031” Award to the Institute of Cosmos Sciences; and by the Generalitat de Catalunya, Grant No. 2021SGR01095.
We also acknowledge J. W. T. Keeble and M. Drissi for relevant discussions.






\bibliography{NQS.bib}


\end{document}